\documentclass[draftcls, 11pt, onecolumn]{IEEEtran}
\usepackage{amsmath,amssymb,cite}
\usepackage[dvips]{graphicx}
\usepackage{url}
\usepackage{hyperref}

\newtheorem{proposition}{Proposition}

\newtheorem{lemma}{Lemma}

\newcommand{\df}{\stackrel{\mbox{\scriptsize def}}{=}}

\newcommand{\rk}{\mathrm{rk}}

\newcommand{\dr}{d_{\mbox{\tiny{R}}}}

\newcommand{\Kr}{K_{\mbox{\tiny{R}}}}

\newcommand{\Jr}{J_{\mbox{\tiny{R}}}}

\begin{document}

\title{Bounds on Covering Codes with the Rank Metric}

\author{Maximilien Gadouleau,~\IEEEmembership{Student Member, IEEE,}
and Zhiyuan Yan,~\IEEEmembership{Senior Member, IEEE}%
\thanks{This work was supported in part by Thales Communications
Inc. and in part by a grant from the Commonwealth of Pennsylvania,
Department of Community and Economic Development, through the
Pennsylvania Infrastructure Technology Alliance (PITA). This work was carried out while M. Gadouleau was at Lehigh University.} %
\thanks{Maximilien Gadouleau is with Crestic, Universit\'e de Reims Champagne-Ardenne, Reims, France. Zhiyuan Yan is with the Department of Electrical and Computer
Engineering, Lehigh University, Bethlehem, PA, 18015 USA (e-mail:
maximilien.gadouleau@univ-reims.fr; yan@lehigh.edu).}}

\maketitle

\thispagestyle{empty}

\begin{abstract}
In this paper, we investigate geometrical properties of the rank metric space and covering properties of rank metric codes. We first establish an analytical expression for the intersection of two balls with rank radii, and
then derive an upper bound on the volume of the union of multiple balls with rank radii. Using these geometrical properties, we derive both upper and lower bounds on the minimum cardinality of a code
with a given rank covering radius. The geometrical properties and bounds proposed in this paper are significant to the design, decoding, and performance analysis of rank metric codes.
\end{abstract}
%\IEEEpeerreviewmaketitle

\begin{keywords}
Error control codes, covering radius, rank metric codes, geometrical properties, intersection number.
\end{keywords}

\section{Introduction}
There is a steady stream of works on rank metric codes due to their
applications to data storage, public-key cryptosystems, and
space-time coding (see \cite{gadouleau_it08_covering,silva_it08} and
the references therein for a comprehensive literature survey), and
interest in rank metric codes is strengthened by their recent
applications in error control for random network coding
\cite{silva_it08}, decribed below. Constant-dimension codes \cite{koetter_it08} are an important class of codes for error and erasure correction in random linear network coding. Using the lifting operation \cite{silva_it08}, rank metric codes
can be readily turned into constant-dimension codes without modifying their distance properties.
It is shown in \cite{koetter_it08} that liftings of rank
metric codes have many advantages compared to general
constant-dimension codes. First, their cardinalities are 
optimal up to a constant; second, the decoding of these codes can be done in either the subspace
metric \cite{koetter_it08} or the rank metric \cite{silva_it08},
and for both scenarios efficient decoding algorithms were proposed.

Despite their significance, many research problems in rank metric
codes remain open. For example, geometrical properties of the rank
metric space are not well studied, and covering properties of rank
metric codes have received little attention with the exception of
our previous work \cite{gadouleau_it08_covering}. The geometrical
properties, such as the intersection number, characterize
fundamental properties of a metric space, and determine the packing
and covering properties of codes defined in the metric space.
Packing and covering properties are significant to the code design,
decoding, and performance analysis of rank metric codes. For
instance, the covering radius can be viewed as a measure of
performance: if the code is used for error correction, then the
covering radius is the maximum weight of a correctable error vector.
The covering radius of a code also gives a straightforward criterion for 
code optimality: if the covering radius of a code is at least
equal to its minimum distance, then more codewords can be added without altering the minimum distance, and hence
the original code is not optimal. Although liftings of rank metric
codes are nearly optimal constant-dimension codes, they have the
largest possible covering radius \cite{gadouleau_it09_cdc}, and
hence they are not optimal constant-dimension codes. This covering property hence is a
crucial result for the design of error control codes for random
linear network coding.

In this paper, we focus on the geometrical properties of the rank
metric space and covering properties of rank metric codes.  We first
establish an analytical expression for the intersection of two balls
with rank radii, and then derive an upper bound on the volume of the
union of multiple balls with rank radii. Both results are novel to
the best of our knowledge. Using these geometrical properties, we
derive novel lower and upper bounds on the minimum cardinality
$\Kr(q^m,n,\rho)$ of a code in $\mathrm{GF}(q^m)^n$ with rank
covering radius $\rho$. The bounds presented in this paper are
obtained based on different approaches from those in
\cite{gadouleau_it08_covering}, and they are the tightest bounds to
the best of our knowledge for many sets of parameter values.

%Due to space limitation, we shall omit some technical proofs, which can be found online at %\href{http://arxiv.org/abs/0809.2968}{http://arxiv.org/abs/0809.2968}.

\section{Preliminaries}
%Viewing the field $\mathrm{GF}(q^m)$ as an $m$-dimensional vector space over $\mathrm{GF}(q)$,
The rank weight of a vector ${\bf x}
\in \mathrm{GF}(q^m)^n$, denoted as $\rk({\bf x})$, is defined to be
the \emph{maximum} number of coordinates in ${\bf x}$ that are linearly
independent over $\mathrm{GF}(q)$. The number of vectors of rank weight $r$
in $\mathrm{GF}(q^m)^n$ is $N_r = {n \brack r} \alpha(m,r)$, where
$\alpha(m,0) \df 1$, $\alpha(m,r) \df \prod_{i=0}^{r-1}(q^m-q^i)$,
and ${n \brack r} \df \alpha(n,r)/\alpha(r,r)$ for $r \geq 1$
\cite{andrews_book76}. The rank distance between ${\bf x}$ and ${\bf
y}$ is defined as $\dr({\bf x}, {\bf y}) = \rk({\bf x} - {\bf y})$.
If a vector ${\bf x}$ is at distance at most $\rho$ from a code $C$,
we say $C$ covers ${\bf x}$ with radius $\rho$. The rank covering
radius $\rho$ of a code $C$ is defined as $\max_{{\bf x} \in
\mathrm{GF}(q^m)^n} \dr({\bf x}, C)$.
%For any basis $B_m$
%of $\mathrm{GF}(q^m)$ over $\mathrm{GF}(q)$, each coordinate of
%${\bf x}$ can be expanded to an $m$-dimensional column vector over
%$\mathrm{GF}(q)$ with respect to $B_m$. The rank weight of ${\bf x}$
%is hence the rank of the $m\times n$ matrix over $\mathrm{GF}(q)$
%obtained by expanding all the coordinates of ${\bf x}$. For all
%${\bf x}, {\bf y}\in \mathrm{GF}(q^m)^n$, it is easily verified that
%$\dr({\bf x},{\bf y})\df \rk({\bf x} - {\bf y})$ is a metric over
%GF$(q^m)^n$ \cite{gabidulin_pit0185}, referred to as the \emph{rank
%metric} henceforth.
%The {\em minimum rank distance} of a code $C$,
%denoted as $d_{\mbox{\tiny R}}$, is simply the minimum rank distance
%over all possible pairs of distinct codewords.
%
%We shall assume the vector space is $\mathrm{GF}(q^m)^n$ henceforth.
%The number of vectors of rank $r$ in $\mathrm{GF}(q^m)^n$ is $N_r =
%{n \brack r} \alpha(m,r)$ \cite{gabidulin_pit0185}, where
%$\alpha(m,0) \df 1$, $\alpha(m,r) \df \prod_{i=0}^{r-1}(q^m-q^i)$,
%and ${n \brack r} \df \alpha(n,r)/\alpha(r,r)$ for $r \geq 1$. The
%volume of a ball with rank radius $r$ is denoted as $v(r) =
%\sum_{i=0}^r N_i$.
%
When $n \leq m$, the minimum rank distance $\dr$ of a code of length $n$ and
cardinality $M$ over $\mathrm{GF}(q^m)$ satisfies $\dr \leq
n-\log_{q^m}M+1$; we refer to this bound as
the Singleton bound for rank metric codes. The equality is attained by a class of linear rank metric codes called maximum rank distance (MRD) codes.

%For all $q$, $1 \leq
%d \leq r \leq n \leq m$, the number of codewords of rank $r$ in an
%$(n, n-d+1, d)$ linear MRD code over $\mathrm{GF}(q^m)$ is given by
%\cite{gabidulin_pit0185}
%\begin{equation}\label{eq:Mdr_def}
%    M(d,r) \df {n \brack r} \sum_{j=d}^r (-1)^{r-j}
%    {r \brack j} q^{r-j \choose 2} \left( q^{m(j-d+1)} - 1\right).
%\end{equation}
%
%The rank covering radius $\rho$ of a code $C$ with length $n$ over
%$\mathrm{GF}(q^m)$ is defined to be the smallest integer $\rho$ such
%that all vectors in the space $\mathrm{GF}(q^m)^n$ are within rank
%distance $\rho$ of some codeword of $C$ \cite{cohen_book97}. It is
%the maximal rank distance from any vector in $\mathrm{GF}(q^m)^n$ to
%the code $C$, that is, $\rho = \max_{{\bf x} \in \mathrm{GF}(q^m)^n}
%\{ \dr({\bf x},C)\}$. The minimum cardinality of a code over
%$\mathrm{GF}(q^m)$ with length $n$ and rank covering radius $\rho$
%is denoted as $\Kr(q^m,n,\rho)$. Upper and lower bounds on
%$\Kr(q^m,n,\rho)$ are derived in \cite{gadouleau_it08_covering}.

\section{Geometrical Properties of balls with rank radii}\label{sec:balls}
%Following \cite{gadouleau_it08_covering}, we denote the volume of a
%ball with rank radius $r$ and the number of codewords with rank $r$
%in an $(n,n-d+1,d)$ linear MRD code as $v(r)$ and $M(d,r)$,
%respectively.
We denote the intersection of two spheres (balls,
respectively) in $\mathrm{GF}(q^m)^n$ with rank radii $u$ and $s$
and centers with distance $w$ as $\Jr(u,s,w)$ ($I(u,s,w)$,
respectively).
\begin{lemma}\label{lemma:J} We have
%The intersection of two balls with rank radii $u$ and $s$ and
%distance between their centers $w$ is given by
\begin{equation}\label{eq:Jr}
    \Jr(u,s,w) = \frac{1}{q^{mn} N_w} \sum_{i=0}^n N_i K_u(i)
    K_s(i) K_w(i),
\end{equation}
where $K_j(i)$ is a $q$-Krawtchouk polynomial
\cite{delsarte_siam76}:
\begin{equation}\label{eq:Kji}
    K_j(i) = \sum_{l=0}^j (-1)^{j-l} q^{lm + {j-l \choose 2}} {n-l \brack
    n-j} {n-i \brack l}.
\end{equation}
\end{lemma}

%\begin{proof}
%This directly follows \cite[Chapter II, Theorem 3.6]{bannai_book83}.
%\end{proof}
Although (\ref{eq:Jr}) is obtained by a direct application of \cite[Chapter II,
Theorem 3.6]{bannai_book83}, we present it formally since it is a
fundamental geometric property of the rank metric space. Since
\begin{equation}\label{eq:I}
    I(u,s,w) = \sum_{i=0}^u \sum_{j=0}^s \Jr(i,j,w),
\end{equation}
(\ref{eq:Jr}) also leads to an analytical expression for $I(u,s,w)$.
In order to simplify notations
%and to be consistent with the
%notations in \cite{gadouleau_it08_covering},
we denote $I(\rho,\rho,d)$ as $I(\rho,d)$.  Both (\ref{eq:Jr}) and (\ref{eq:I}) are
instrumental in our later derivations.

We denote the volume of a ball with rank radius $\rho$ as $v(\rho)$,
and we now derive a bound on the volume of the union of any $K$
balls with radius $\rho$, which will be instrumental in
Section~\ref{sec:bounds}.

\begin{lemma}\label{lemma:B}
The volume of the union of \emph{any} $K$ balls with rank radius $\rho$ is at
most
\begin{align}
    \nonumber
    B(K) =&  v(\rho) + \sum_{a=1}^l (q^{am} - q^{(a-1)m})
    [v(\rho) - I(\rho,n-a+1)]\\
    & + (K - q^{lm}) [v(\rho) - I(\rho,n-l)],
    %\\ \label{eq:B} ,
\end{align}
where $l = \lfloor \log_{q^m} K \rfloor$.
\end{lemma}

\begin{proof}
Let $\{ {\bf v}_i \}_{i=0}^{K-1}$ denote the centers of $K$ balls
with rank radius $\rho$ and let $V_j = \{ {\bf v}_i \}_{i=0}^{j-1}$
for $1 \leq j \leq K$. The centers are labeled such that $\dr({\bf
v}_j, V_j)$ is non-increasing for $j
> 0$. For $1 \leq a \leq l$ and $q^{m(a-1)} \leq j < q^{ma}$, we
have $\dr({\bf v}_j, V_j) = \dr(V_{j+1}) \leq n-a+1$ by the
Singleton bound. The center ${\bf v}_j$ hence covers at most
$v(\rho) - I(\rho, n-a+1)$ vectors that are not previously covered
by $V_j$.
%The number of vectors covered by $V_K$ is thus at most $B(K)$.
\end{proof}

\section{Bounds on covering codes with the rank
metric}\label{sec:bounds} We consider covering codes in
$\mathrm{GF}(q^m)^n$, and without loss of generality we assume
$n\leq m$ due to transposition \cite{gadouleau_it08_covering}.
We first derive a lower bound on $\Kr(q^m,n,\rho)$ based on the
linear inequalities satisfied by covering codes.

\begin{proposition}\label{prop:T}
For $0 \leq \delta \leq \rho$, let $T_\delta = \min \sum_{i=0}^n
A_i(\delta)$, where the minimum is taken over all integer sequences
$\{A_i(\delta)\}$ which satisfy $A_i(\delta) = 0$ for $0 \leq i \leq
\delta-1$, $A_\delta(\delta) \geq 1$, $0 \leq A_i(\delta) \leq N_i$
for $\delta+1 \leq i \leq n$, and $\sum_{i=0}^n A_i(\delta)
\sum_{s=0}^\rho \Jr(r,s,i) \geq N_r$ for $0 \leq r \leq n$. Then
$\Kr(q^m,n,\rho) \geq \max_{0 \leq \delta \leq \rho} T_\delta$.
\end{proposition}

\begin{proof}
Let $C$ be a code with covering radius $\rho$. For any ${\bf u} \in
\mathrm{GF}(q^m)^n$ at distance $0 \leq \delta \leq \rho$ from $C$,
let $A_i(\delta)$ denote the number of codewords at distance $i$
from ${\bf u}$. Then $\sum_{i=0}^n A_i(\delta) = |C|$ and we easily
obtain $A_i(\delta) = 0$ for $0 \leq i \leq \delta-1$,
$A_\delta(\delta) \geq 1$, and $0 \leq A_i(\delta) \leq N_i$ for
$\delta+1 \leq i \leq n$. Also, for $0 \leq r \leq n$, all the
vectors at distance $r$ from ${\bf u}$ are covered, hence
$\sum_{i=0}^n A_i(\delta) \sum_{s=0}^\rho \Jr(r,s,i) \geq N_r$.
\end{proof}

We note that the notation $A_i(\delta)$ is used above since the constraints on the sequence depend on $\delta$. Since $T_\delta$ is the solution of an integer linear programming,
it is computationally infeasible to determine $T_\delta$ for very
large parameter values.

%\subsection{An upper bound based on a greedy
%algorithm}\label{sec:domination}

%A dominating set of a graph is a set of vertices whose neighborhood
%is the whole graph, and the domination number of a graph is the
%smallest cardinality of a dominating set \cite{haynes_book98}. Let
%$G$ be the graph where the vertex set is $\mathrm{GF}(q^m)^n$ and
%two vertices are adjacent if and only if their rank distance is at
%most $\rho$. Hence, a dominating set of $G$ is a code with rank
%covering radius $\leq \rho$ and the domination number of $G$ is
%equal to $\Kr(q^m,n,\rho)$. An upper bound on the domination number
%of any graph based on a greedy algorithm is derived in
%\cite{clark_ejc97}.

We now derive an upper bound on $\Kr(q^m,n,\rho)$ by providing a
nontrivial refinement of the greedy algorithm in \cite{clark_ejc97}.

\begin{lemma}\label{lemma:domination}
Let $C$ be a code which covers at least $q^{mn}-u$ vectors in
$\mathrm{GF}(q^m)^n$ with radius $\rho$. Then for any $k \geq |C|$,
there exists a code with cardinality $k$ which covers at least
$q^{mn}-u_k$ vectors, where $u_{|C|} = u$ and for $k \geq |C|$
\begin{equation}\label{eq:u}
    u_{k+1} = u_k - \left\lceil \frac{u_k v(\rho)}{\min\{q^{mn} - k, B(u_k)\}}
    \right\rceil.
\end{equation}
Thus $\Kr(q^m,n,\rho) \leq \min\{k : u_k = 0\}$.
\end{lemma}

\begin{proof}
The proof is by induction on $k$. By hypothesis, $C$ is a code with
cardinality $|C|$ which leaves $u_{|C|}$ vectors uncovered. Suppose
there exists a code with cardinality $k$ which leaves $t_k \leq u_k$
vectors uncovered, and denote the set of uncovered vectors as $T_k$.
Let $G$ be a graph where the vertex set is $\mathrm{GF}(q^m)^n$
and two vertices are adjacent if and only if their rank distance is
at most $\rho$. Let ${\bf A}$ be the adjacency matrix of $G$ and
${\bf A}_k$ be the $t_k$ columns of ${\bf A}$ corresponding to
$T_k$. There are $t_k v(\rho)$ ones in ${\bf A}_k$, distributed
across $|N(T_k)|$ rows, where $N(T_k)$ is the neighborhood of $T_k$.
By construction, $N(T_k)$ does not contain any codeword, hence
$|N(T_k)| \leq q^{mn} - k$. Also, by Lemma~\ref{lemma:B}, $|N(T_k)|
\leq B(t_k) \leq B(u_k)$. Thus $|N(T_k)| \leq \min\{q^{mn} - k,
B(u_k)\}$ and there exists a row with at least $
    \left\lceil \frac{t_k v(\rho)}{\min\{q^{mn} - k, B(u_k)\}}
    \right\rceil
$ ones in ${\bf A}_k$. Adding the vector corresponding to this row
to the code, we obtain a code with cardinality $k+1$ which leaves at
most $t_k - \left\lceil \frac{t_k v(\rho)}{\min\{q^{mn} - k,
B(u_k)\}} \right\rceil \leq u_{k+1}$ vectors uncovered.
\end{proof}

The upper bound in Lemma~\ref{lemma:domination} is nontrivial since
the sequence $\{u_k\}$ is monotonically
decreasing until it reaches $0$ for $k \leq q^{mn}$. Since the sequence $\{u_k\}$
depends on $C$ only through the number of vectors covered by $C$,
to obtain a tighter bound on $\Kr(q^m,n,\rho)$ based on
Lemma~\ref{lemma:domination}, it is necessary to find a code $C$
which leaves a smaller number of vectors uncovered. We consider two
choices for $C$ below. Since any codeword can cover at most
$v(\rho)$ vectors uncovered by the other codewords, we have $u \geq
q^{mn} - v(\rho) |C|$. In the first choice, we consider the largest
code $C_0$ that achieves $u = q^{mn} - v(\rho) |C_0|$, which is an
$(n,n-a,a+1)$ linear MRD code with $a = \min\{n,2\rho\}$.

An alternative choice would be to select a supercode of $C_0$,
that is, an MRD code with a larger dimension. Since there may be
intersection between balls of radius $\rho$ around codewords, we now
derive a lower bound on the number of vectors covered by only one
codeword in an MRD code. Let us denote the number of codewords with rank $r$ in an
$(n,n-d+1,d)$ linear MRD code as $M(d,r)$.

\begin{lemma}\label{lemma:lambda}
Let the vectors ${\bf c}_j \in \mathrm{GF}(q^m)^n$ be sorted such
that $\{{\bf c}_j\}_{j=0}^{q^{m(n-l+1)}-1}$ is an $(n,n-l+1,l)$ MRD
code for $1 \leq l \leq n$. For all $1 \leq k \leq q^{mn}-1$, we
denote $\{{\bf c}_j\}_{j=0}^{k-1}$ as  $C_k$ and its minimum rank
distance is given by $d_k = n - \lceil \log_{q^m} (k+1) \rceil + 1$.
Then for $d_k \geq 2\rho + 1$, ${\bf c}_k$ covers $v(\rho)$ vectors
not covered by $C_k$. For $d_k \leq 2\rho$, ${\bf c}_{k}$ covers at
least $v(\rho) - \sum_{l=d_k}^a \mu_l I(\rho,l)$
%\begin{equation}
%    \lambda_{k+1} =  v(\rho) - \sum_{l=d_k}^a \mu_l I(\rho,l)
%\end{equation}
vectors not covered by $C_k$, where $a = \min\{n,2\rho\}$, $\mu_{d_k} =
\min\{M(d_k,d_k), k\}$, and $\mu_l = \min\Big\{ M(d_k,l), k -
\sum_{j=d_k}^{l-1} \mu_j \Big\}$ for $l > d_k$.
\end{lemma}

\begin{proof}
The case $d_k \geq 2\rho + 1$ is straightforward. We assume $d_k \leq
2\rho$ henceforth.
%It is clear from the minimum distance of
%$C_{k+1}$ that any codeword covers at least $v(\lfloor \frac{d-1}{2}
%\rfloor)$ vectors which are not covered by any other codeword.
We denote the number of codewords in $C_k$ at distance $l$ ($0 \leq
l \leq n$) from ${\bf c}_k$ as $M_l^{(k)}$. Since a codeword in
$C_k$ at distance $l$ from ${\bf c}_k$ covers exactly $I(\rho,l)$
vectors also covered by ${\bf c}_k$, ${\bf c}_k$ covers at least
$v(\rho) - \sum_{l=d_k}^a M_l^{(k)} I(\rho,l)$ vectors that are not
covered by $C_k$. Since the value of $M_l^{(k)}$ is unknown, we give
a lower bound on the quantity above. Since $I(\rho,l)$ is a
non-increasing function of $l$ \cite{gadouleau_it08_covering}, the
sequence $\{\mu_l\}$ as defined above minimizes the value of
$v(\rho) - \sum_{l=d_k}^a M_l^{(k)} I(\rho,l)$ under the constraints
$0 \leq M_l^{(k)} \leq M(d_k,l)$ and $\sum_{l=d_k}^n M_l^{(k)} = k$.
%Since $I(\rho,l)$ is a non-increasing function of
%$l$ \cite{gadouleau_it08_covering}, it is easily shown that the
%minimum is obtained for the sequence $\{\mu_l\}$.
Thus, ${\bf c}_k$ covers at least $v(\rho) - \sum_{l=d_k}^a \mu_l I(\rho,l)$ vectors not
covered by $C_k$.
\end{proof}

\begin{table*}[!htp]
\begin{center}
\begin{tabular}{|c|c|cccccc|}
    \hline
    $m$ & $n$ & $\rho=1$ & $\rho=2$ & $\rho=3$ & $\rho=4$ &
    $\rho=5$ & $\rho=6$\\
    \hline
    2 & 2 & 3 & 1 & & & &\\
    \hline
    3 & 2 & 4 & 1 & & & &\\
      & 3 & 11-16 & 4 & 1 & & &\\
    \hline
    4 & 2 & 7-8 & 1 & & & &\\
      & 3 & 40-64 & 4-7 & 1 & & &\\
      & 4 & 293-722 & 10-48 & 3-{\bf 5 J} & 1 & &\\
    \hline
    5 & 2 & 12-16 & 1 & & & &\\
      & 3 & 154-256 & 6-8 & 1 & & &\\
      & 4 & 2267-4096 & 33-256 & 4-8 & 1 & &\\
      & 5 & 34894-$2^{17}$ & 233-{\bf 2773 I} & {\bf h 10}-32 & 3-{\bf 6 J} & 1&\\
    \hline
    6 & 2 & 23-32 & 1 & & & &\\
      & 3 & 601-1024 & 11-16 & 1 & &  &\\
      & 4 & 17822-$2^{15}$ & 124-256 & 6-16 & 1 &  &\\
      & 5 & 550395-$2^{20}$ & 1770-$2^{14}$ & 31-256 & {\bf i 4}-16 & 1 &\\
      & 6 & 17318410-$2^{26}$ & 27065-{\bf 401784 I} & 214-{\bf 4092 I} & 9-{\bf 154 J} &
      3-{\bf 7 J} & 1\\
    \hline
    7 & 2 & 44-64 & 1 & & & &\\
      & 3 & 2372-4096 & 20-32 & 1 & & & \\
      & 4 & 141231-$2^{18}$ & 484-1024 & {\bf i 10}-16 & 1 & & \\
      & 5 & 8735289-$2^{24}$ & 13835-$2^{15}$ & {\bf i 112}-1024 & {\bf i 6}-16 & 1 &\\
      & 6 & 549829402-$2^{30}$ & 42229-$2^{22}$ & {\bf i 1585}-$2^{15}$ & 29-{\bf 708 I} &
      {\bf i 4}-16 & 1\\
      & 7 & 34901004402-$2^{37}$ & 13205450-{\bf 233549482 I} & {\bf i 23979}-{\bf 573590 I} & 203-{\bf 5686 I} &
      {\bf h 9}-{\bf 211 J} & 3-{\bf 8 J}\\
    \hline
\end{tabular}
\caption{Bounds on $\Kr(2^m,n,\rho)$, for $2 \leq m \leq 7$, $2 \leq
n \leq m$, and $1 \leq \rho \leq 6$.}\label{table:bounds}
\end{center}
\end{table*}

\begin{proposition}\label{prop:bound_domination}
Let $a = \min\{n,2\rho\}$, $K_0 = q^{m(n-a)}$, and $u_{K_0} = q^{mn}
- q^{m(n-a)}v(\rho)$. Consider two sequences $\{h_k\}$ and
$\{u'_k\}$, both of which are upper bounds on the number of vectors
yet to be covered by a code of cardinality $k$, given by $h_{K_0} =
u'_{K_0} = u_{K_0}$ and for $k \geq K_0$
\begin{eqnarray}
    \nonumber
    h_{k+1} &=& h_k - v(\rho) + \sum_{l=d}^a \mu_l I(\rho,l),\\
    \nonumber
    u'_{k+1} &=& \min \left\{h_{k+1}, u'_k - \left\lceil \frac{u'_k v(\rho)}{\min\{q^{mn} - k, B(u'_k)\}}
    \right\rceil\right\}.
\end{eqnarray}
Then, $\Kr(q^m,n,\rho) \leq \min\{k : u'_k = 0\}$.
\end{proposition}

\begin{proof}
Let $C_0$ be an $(n,n-a,a+1)$ MRD code over $\mathrm{GF}(q^m)$.
$C_0$ has cardinality $K_0$ and covers $q^{m(n-a)} v(\rho) = q^{mn}
- u_{K_0}$ vectors in $\mathrm{GF}(q^m)^n$. By
Lemma~\ref{lemma:lambda}, adding $k-K_0$ codewords of an MRD code
which properly contains $C_0$ leads to a code with cardinality $k$
which covers at least $q^{mn} - h_k$ vectors. On the other hand,
$u'_k$ is the upper bound on the number of vectors yet to be covered
by a code of cardinality $k$ that is obtained recursively either by
applying the greedy approach in the proof of
Lemma~\ref{lemma:domination} or by adding more codewords based on
the MRD codes as described in the proof of Lemma~\ref{lemma:lambda}.
The recursion of $u'_k$ follows from this construction. Since this
recursion is constructive, there exists a code with cardinality $k$
which leaves at most $u'_k$ vectors uncovered, and hence
$\Kr(q^m,n,\rho) \leq \min\{k : u'_k = 0\}$.
\end{proof}

%\subsection{New covering codes with the rank
%metric}\label{sec:hamming}

We now construct a new class of covering
codes with the rank metric. We assume all the vectors in
$\mathrm{GF}(q^m)^n$ are expanded with respect to a given basis of
$\mathrm{GF}(q^m)$ to $m \times n$ matrices over $\mathrm{GF}(q)$.
First, for any positive integer $k$, we denote $S_k \df \{0, 1,
\ldots, k-1\}$. For ${\bf V} \in \mathrm{GF}(q)^{m \times n}$, $I
\subseteq S_m$, and $J \subseteq S_n$, we denote ${\bf V}(I,J) =
(v_{i,j})_{i \in I, j \in J}$, where the indexes are all sorted
increasingly. Consider the code $C$ which consists of all matrices ${\bf C} \in \mathrm{GF}(q)^{m
\times n}$ with ${\bf C}(S_\rho, S_n) = {\bf 0}$ and with at most
$n-\rho$ nonzero columns.
%, and ${\bf V}(i,J) = {\bf V}(\{i\},J)$.
\begin{proposition}\label{prop:bound_hamming}
$C$ has covering radius $\rho$ and
$
%\begin{equation}\label{eq:bound_hamming}
    \Kr(q^m,n,\rho) \leq |C| =  \sum_{i=0}^{n-\rho} {n \choose
    i} (q^{m-\rho} - 1)^i.
%\end{equation}
$
\end{proposition}
\begin{proof}
First, the cardinality of $C$ is given by the number of vectors in
$\mathrm{GF}(q^{m-\rho})^n$ with Hamming weight at most $n-\rho$.
%, i.e., $|C| = \sum_{i=0}^{n-\rho} {n \choose i} (q^{m-\rho} - 1)^i.$
It remains to prove that $C$ has rank covering radius $\rho$.
Suppose ${\bf V} \in \mathrm{GF}(q)^{m \times n}$ has $\rk({\bf
V}(S_\rho,S_n)) = r$ ($0\leq r \leq \rho$),  then there exist $I \subseteq S_\rho$ and $J \subseteq S_n$ such that
$|I| = |J| = \rk({\bf V}(I,J))  = \rk({\bf V}(S_m,J))  = \rk({\bf V}(I,S_n)) = r$. For $\rho \leq i \leq m-1$,
let ${\bf l}_i = {\bf V}(\{i\},J) {\bf V}(I,J)^{-1} \in
\mathrm{GF}(q)^{1 \times r}$. Define ${\bf U}(S_\rho,S_n) = {\bf
V}(S_\rho,S_n)$ and ${\bf U}(\{i\},S_n) = {\bf l}_i {\bf V}(I,S_n)$ for
all $\rho \leq i \leq m-1$. All the rows of ${\bf U}$ are in the row
span of ${\bf V}(I,S_n)$, therefore $\rk({\bf U}) = r$. Also, ${\bf
U}(\{i\},J) = {\bf l}_i {\bf V}(I,J)= {\bf V}(\{i\},J)$ for $\rho \leq i
\leq m-1$ and hence ${\bf U}(S_m,J) = {\bf V}(S_m,J)$.
Thus there exists ${\bf U} \in
\mathrm{GF}(q)^{m \times n}$ such that $\rk({\bf U}) = r$, ${\bf
U}(S_\rho,S_n) = {\bf V}(S_\rho,S_n)$, and ${\bf U}(S_m,J) = {\bf
V}(S_m,J)$. We also construct
${\bf U}' \in \mathrm{GF}(q)^{m \times n}$ by setting ${\bf
U}'(S_m,J') = {\bf V}(S_m,J')$ and ${\bf U}'(S_m, S_n \backslash J')
= {\bf U}(S_m, S_n \backslash J')$, where $|J'| = \rho - r$ and $J
\cap J' = \emptyset$. For ${\bf C} = {\bf V} - {\bf U}'$, ${\bf C}(S_\rho,S_n) = {\bf 0}$ and ${\bf C}(S_m, J
\cup J') = {\bf 0}$. Therefore, ${\bf C} \in C$ and $\dr({\bf V},
{\bf C}) = \rk({\bf U}') \leq \rk({\bf U}) + |J'| = \rho$.
\end{proof}

%\begin{lemma}\label{lemma:U}
%Suppose ${\bf V} \in \mathrm{GF}(q)^{m \times n}$ has $\rk({\bf
%V}(S_\rho,S_n)) = r$. Then there exists ${\bf U} \in
%\mathrm{GF}(q)^{m \times n}$ such that $\rk({\bf U}) = r$, ${\bf
%U}(S_\rho,S_n) = {\bf V}(S_\rho,S_n)$, and ${\bf U}(S_m,J) = {\bf
%V}(S_m,J)$ for some $J \subseteq S_n$ with $|J| = r$.
%\end{lemma}
%
%\begin{proof}
%
%\end{proof}

%Using Lemma~\ref{lemma:U}, we construct a new class of covering
%codes with the rank metric.

%It is remarkable that a bound on the sphere covering problem in the
%rank metric is given by the volume of a ball with Hamming radius in
%another space.

%Because the RHS of (\ref{eq:bound_hamming}) increases rapidly with
%$n-\rho$ and $m-\rho$, the bound in
%Proposition~\ref{prop:bound_hamming} is tight for $\rho$ and $n$
%close to $m$ only.

%\subsection{Table}

In order to illustrate the improvement due to the bounds in this paper,
similar to \cite{gadouleau_it08_covering}, we provide the tightest
lower and upper bounds on $\Kr(2^m,n,\rho)$ for $2 \leq m \leq 7$, $2 \leq n \leq
m$, and $1 \leq \rho \leq 6$ in Table~\ref{table:bounds}. The
improved entries in Table~\ref{table:bounds} due to the results in
this paper are boldface, and are associated with letters indicating the
sources.
%The letters in Table~\ref{table:bounds} indicate their sources.
The lower bound on $\Kr(q^m,n,\rho)$ in \cite[Proposition
8]{gadouleau_it08_covering} is based on $I(\rho,d)$. However, due to
the lack of analytical expression for $I(\rho,d)$, in
\cite{gadouleau_it08_covering} we were able to compute $I(\rho,d)$
only for small values, using exhaustive search. Using
(\ref{eq:Jr})-(\ref{eq:I}), we calculate $I(u,s,w)$ and hence the
bound in \cite[Proposition 8]{gadouleau_it08_covering} for any set
of parameter values, and the improved entries for this reason are
associated with the lower case letter i. The lower case letter h and
the upper case letters I and J correspond to improvements due to
Propositions~\ref{prop:T},~\ref{prop:bound_domination},
and~\ref{prop:bound_hamming}, respectively. The unmarked entries in Table~\ref{table:bounds} are the same as those in
\cite{gadouleau_it08_covering}.

\bibliographystyle{IEEEtran}
\bibliography{gpt}

\end{document}